\documentclass[conference]{IEEEtran}
\IEEEoverridecommandlockouts

\usepackage{cite}
\usepackage{amsmath,amssymb,amsfonts}
\usepackage{algorithm}      
\usepackage{algorithmic}
\usepackage{graphicx}
\usepackage{textcomp}
\usepackage{xcolor}
\usepackage{float}
\usepackage{bm}
\usepackage{soul}
\usepackage{subcaption}
\usepackage{stfloats}
\usepackage{siunitx}

\graphicspath{{graph/}}
\DeclareGraphicsExtensions{.pdf,.png,.jpg,.jpeg}

\def\BibTeX{{\rm B\kern-.05em{\sc i\kern-.025em b}\kern-.08em
    T\kern-.1667em\lower.7ex\hbox{E}\kern-.125emX}}

\begin{document}

\title{Robust Covariance-Based DoA Estimation under Weather-Induced Distortion}

\author{
\IEEEauthorblockN{
    Chenyang Yan\IEEEauthorrefmark{1}
    Geert Leus \IEEEauthorrefmark{2}
    Mats Bengtsson \IEEEauthorrefmark{1}
    }
\IEEEauthorblockA{
\IEEEauthorrefmark{1}KTH Royal Institute of Technology
\IEEEauthorrefmark{2}Delft University of Technology
}
\thanks{This work has been funded by Trafikverket, through the FutuRe project, from the Europe's Rail Joint Undertaking under the
European Union’s Horizon Europe research and innovation programme, grant agreement No. 101101962. This publication reflects only the author’s view and the EuRail JU is not responsible for any use that may be made of the
information it contains.}
}

\maketitle

\begin{abstract}
We investigate robust direction-of-arrival (DoA) estimation for sensor arrays operating in adverse weather conditions, where weather-induced distortions degrade estimation accuracy. Building on a physics-based $S$-matrix model established in prior work, we adopt a statistical characterization of random phase and amplitude distortions caused by multiple scattering in rain. Based on this model, we develop a measurement framework for uniform linear arrays (ULAs) that explicitly incorporates such distortions. To mitigate their impact, we exploit the Hermitian Toeplitz (HT) structure of the covariance matrix to reduce the number of parameters to be estimated. We then apply a generalized least squares (GLS) approach for calibration. Simulation results show that the proposed method effectively suppresses rain-induced distortions, improves DoA estimation accuracy, and enhances radar sensing performance in challenging weather conditions.
\end{abstract}


\section{Introduction}
\label{sec:1}
Direction-of-arrival (DoA) estimation, also known as direction finding, involves determining the spatial spectra of impinging signals and is a fundamental problem in sensor array signal processing with wide-ranging engineering applications~\cite{chung2014doa}. A key application of DoA estimation is in radar detection and imaging. For instance, radar sensors are integrated into autonomous vehicles for collision avoidance and autonomous driving, and multiple-input multiple-output (MIMO) radar arrays can be deployed at level crossings for obstacle detection~\cite{waldschmidt2021automotive,narayanan2011railway}. The accuracy of these applications heavily depends on DoA estimation and can be significantly degraded under adverse weather conditions, such as rain, fog, snow, hail, dust, and sand~\cite{zang2019impact,kawaguchi2024experimental}.

Adverse weather primarily impacts radar sensors through attenuation and backscattering: attenuation reduces received signal power, while backscattering introduces interference. Research by the Ballistic Research Laboratory highlighted rain as the most impactful weather condition for radar, due to raindrop sizes comparable to radar wavelengths~\cite{Ballistic}. Additionally, mathematical models for snow and mist closely resemble those for rain~\cite{pozhidaev2010estimation}. Therefore, this study focuses specifically on rainy conditions, although the methods developed here are broadly applicable.

While most existing studies rely on attenuation coefficients or signal-to-interference-plus-noise ratios (SINR) without explicitly modeling the statistical impact of weather on wave phase distortion and amplitude variations, recent research introduces a physics-based propagation model for radar signals in random media using the $S$-matrix method~\cite{yektakhah2024model,yektakhah2023physics}. This method partitions the random medium into multiple slabs, each modeled as a multi-port network of electric fields, thereby preserving detailed phase information critical for accurate DoA estimation.

In this work, we adopt the $S$-matrix-based model for phase and amplitude distortions under rainy conditions, as described in Section~\ref{sec:2}. We introduce an array model and associated measurement framework incorporating weather-induced distortions in Section~\ref{sec:3}. Subsequently, we detail DoA estimation methods and calibration algorithm in Sections~\ref{sec:4} and~\ref{sec:5}. The performance of this method is validated through simulations in Section~\ref{sec:6}, followed by conclusions in Section~\ref{sec:7}.

\section{Distortion Model under Adverse Weather Conditions}\label{sec:2}
To model the distortion effects induced by rain on radar wave propagation, we adopt the statistical characterization presented in~\cite{yektakhah2024model}. In this model, consider two points P1 and P2 located on the same wavefront, at equal range from the source, separated by a distance $d$. The normalized fluctuating electric fields at these points are modeled as zero-mean, jointly circularly symmetric complex Gaussian random variables, denoted by $b_1$ and $b_2$. Their second-order statistics are given by:
\begin{align}
    E\left[|b_1|^2\right] &= E\left[|b_2|^2\right] = 2\lambda_{11}, \\
    E\left[b_1 b_2^*\right] &= 2\lambda_{13}.
\end{align}
which is equivalent to the assumptions made in~\cite{yektakhah2024model}.
The notation $\lambda_{11}$ and $\lambda_{13}$ is consistent with~\cite{yektakhah2024model}, and $\lambda_{13}$ is real-valued.

Under this setup, and following the methodology in~\cite{yektakhah2024model,sarabandi1992derivation}, simplified joint probability density functions (pdfs) for the phase difference $\phi = \angle\{b_1\} - \angle\{b_2\}$ and the magnitude ratio $r = |b_1 / b_2|$ between two points separated by a distance $d$, can be derived as equations (9a) and (9b) in \cite{yektakhah2024model}. These pdfs depend on a single real-valued parameter $\alpha = \lambda_{13} / \lambda_{11}$, which is determined by the rain rate, the distance from the source, the wavefront separation $d$, and the radar frequency.

An empirical model for the parameter $\alpha$ presented in~\cite{yektakhah2024model} is expressed as:
\begin{equation}
\alpha = \exp\left(-a_1 \left(\frac{R}{a_2 R + 1}\right) \left(\frac{\frac{d}{\lambda_0}}{a_3 \frac{d}{\lambda_0} + 1}\right)\right),
\label{empirical}
\end{equation}
where $R$ is the range, $\lambda_0$ is the wavelength, and parameters $a_1$, $a_2$, and $a_3$ are related to the rain rate and the wave frequency, which can be obtained from~\cite[Table~II]{yektakhah2024model}. This model is valid for $R \leq 500 \ \mathrm{m}$ and $0.1 \leq \frac{d}{\lambda_0} \leq 8$ according to~\cite{yektakhah2024model}.

\begin{figure}[htbp]
\centerline{\includegraphics[width=\columnwidth]{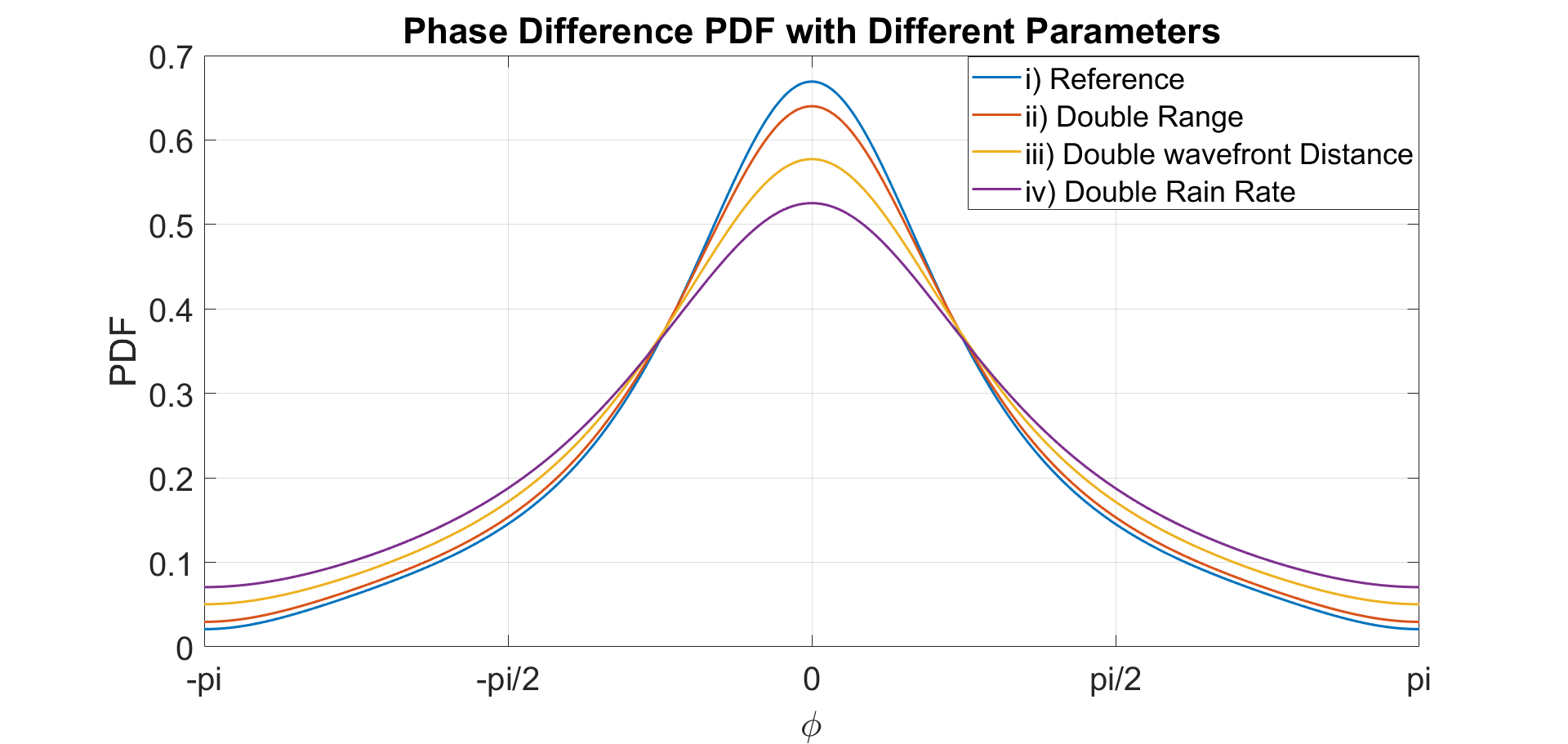}}
\caption{Phase difference pdf with different parameters}
\label{pdf_demo}
\end{figure}

\begin{figure}[htbp]
\centerline{\includegraphics[width=\columnwidth]{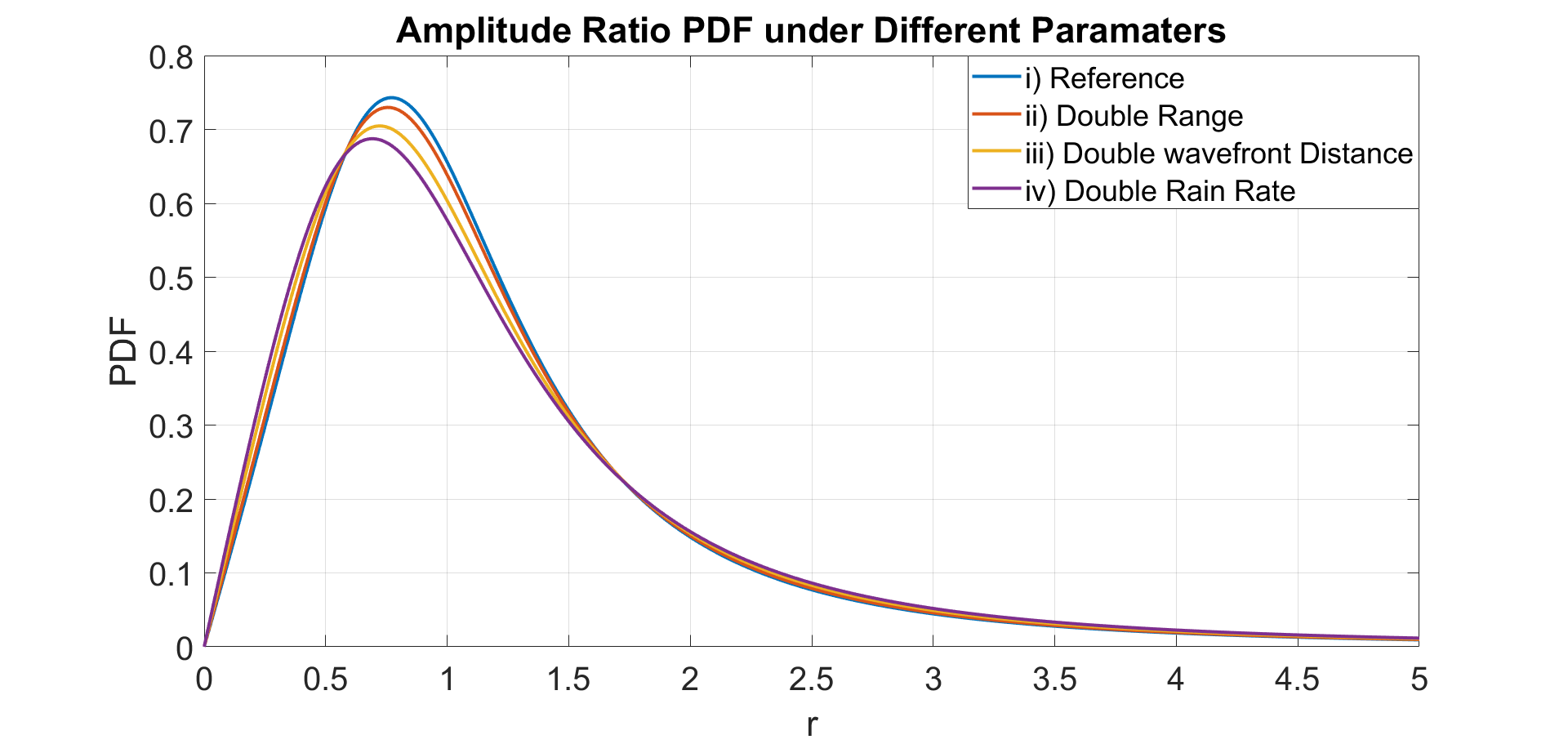}}
\caption{Magnitude ratio pdf with different parameters}
\label{ratio}
\end{figure}

Figures~\ref{pdf_demo} and~\ref{ratio} show the probability density functions (pdfs) of the phase difference and magnitude ratio under varying parameters. Four cases are evaluated, with the corresponding parameter settings summarized in Table~\ref{parameter_table}. The reference case (i) uses $d = 4 \lambda_0$, $R = 200\ \mathrm{m}$, and a rain rate of $25\ \mathrm{mm/hr}$. In case (ii), the range is doubled to $400\ \mathrm{m}$; in case (iii), the wavefront separation is doubled to $8 \lambda_0$; and in case (iv), the rain rate is increased to $50\ \mathrm{mm/hr}$, while the other parameters remain unchanged.

\begin{table}[htbp]
\centering
\caption{Parameter settings and corresponding $\alpha$ values for Figures~\ref{pdf_demo} and~\ref{ratio}}
\begin{tabular}{c|c|c|c|c}
\hline
Case & $d$ $(\lambda_0)$ & $R$ (m) & Rain rate (mm/hr) & $\alpha$ \\
\hline
(i) & 4 & 200 & 25 & 0.6470 \\[2pt]
(ii) & 4 & 400 & 25 & 0.6217 \\[2pt]
(iii) & 8 & 200 & 25 & 0.5598 \\[2pt]
(iv) & 4 & 200 & 50 & 0.4994 \\[2pt]
\hline
\end{tabular}
\label{parameter_table}
\end{table}

As observed, increasing the range, wavefront separation, or rain rate leads to stronger phase and amplitude fluctuations. Specifically, the peak of the phase difference pdf becomes less concentrated around $0^\circ$, and the peak of the magnitude ratio pdf shifts further from 1. Additionally, the phase difference pdfs remain symmetric about the y-axis.

\section{Measurement Model}\label{sec:3}
\begin{figure}[htbp]
\centerline{\includegraphics[width=\columnwidth]{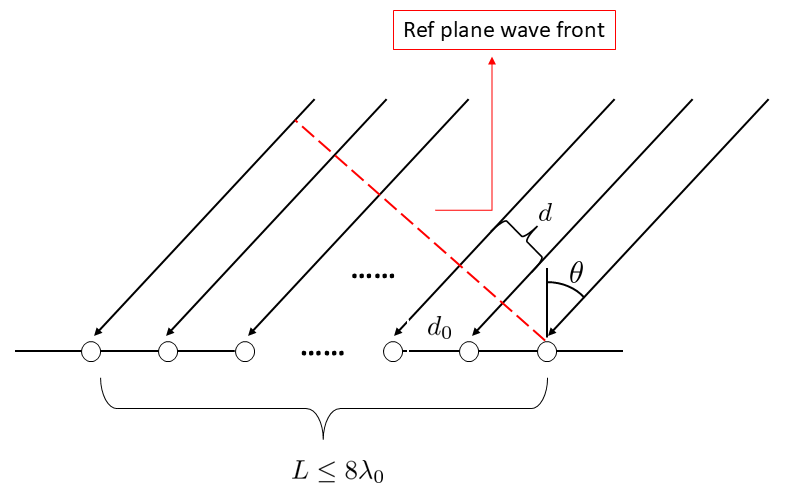}}
\caption{Illustration of the array measurement model with the rain phase distortion.}
\label{array_model}
\end{figure}

We consider a uniform linear array (ULA) consisting of $M$ antennas with spacing $d_0$ between adjacent elements. As illustrated in Figure~\ref{array_model}, we assume that the electric field fluctuations described in Section~\ref{sec:2} occur across the plane wave as it reaches the first antenna encountered during propagation. This plane wave is referred to as the \emph{reference plane wavefront}. Furthermore, fluctuations beyond this reference plane are assumed negligible. To ensure the validity of the empirical model presented in~\eqref{empirical}, the total array aperture should not exceed $8\lambda_0$, corresponding to a maximum of 17 antennas with half-wavelength spacing.

If a narrowband signal $s(t)$ with frequency $f$ is transmitted, the received signal at the $m$-th antenna of the ULA is given by:
\begin{equation}
\begin{split}
    y_m(t) &= \gamma s(t) a_m(\theta) b_m(t) +  n_m(t)\\
    &= \gamma s(t) |b_m| e^{j\left[2\pi m \Delta \sin(\theta) + \phi_m(t)\right]} + n_m(t),
 \end{split}
\end{equation}
where $a_m(\theta) = e^{j2\pi m \Delta \sin(\theta)}$ is the $m$-th entry of the array steering vector corresponding to direction $\theta$, and $b_m(t)$ is the complex distortion introduced by the rain medium, modeled as $b_m(t) = |b_m(t)| e^{j\phi_m(t)}$.  $\gamma$ represents the complex attenuation related to the propagation distance from the source. Additionally, $\Delta = d_0/\lambda_0$ denotes the normalized spacing between adjacent antennas in wavelengths, $\lambda_0 = c/f$ with $c$ as the speed of light, and $\theta$ is the direction of the source. Finally, $ n_m(t)$ is the additive noise at the $m$-th antenna.

The received signal at the array elements can now be expressed as the standard ULA received model element-wise multiplied by a time-varying distortion vector plus some additive noise $\mathbf{n}(t)$:
\begin{equation}
\mathbf{y}(t) =
\begin{bmatrix}
1 \\
e^{j2\pi\Delta \sin \theta} \\
\vdots \\
e^{j2(M-1)\pi\Delta \sin \theta}
\end{bmatrix}
\odot
\begin{bmatrix}
b_1(t) \\
b_2(t)\\
\vdots \\
b_M(t)
\end{bmatrix}
s(t)+\mathbf{n}(t),
\label{full receive model}
\end{equation}
where $\odot$ denotes the Hadamard product (or element-wise product), and the complex propagation attenuation factor $\gamma$ has been absorbed into each $b_m(t)$.

Define the array response vector of the ULA and the distortion vector as $\mathbf{a}(\theta)$ and $\mathbf{b}(t)$ respectively, given by:
\begin{equation}
\mathbf{a}(\theta) =
\begin{bmatrix}
1 \\
e^{j2\pi\Delta \sin(\theta)} \\
\vdots \\
e^{j2(M-1)\pi\Delta \sin(\theta)}
\end{bmatrix},
\end{equation}
\begin{equation}
\mathbf{b}(t) =
\begin{bmatrix}
b_1(t) \\
b_2(t)\\
\vdots \\
b_M(t)
\end{bmatrix}.
\end{equation}
As a result, (\ref{full receive model}) can be expressed in a more concise form as:
\begin{equation}
    \mathbf{y}(t)=\left[\mathbf{a}(\theta) \odot \mathbf{b}(t) \right] s(t)+ \mathbf{n}(t).
\label{concise receive model}
\end{equation}

Assuming that both \( s(t) \) and \( \mathbf{n}(t) \) are i.i.d., circularly symmetric complex processes with zero mean and covariance matrices \( \mathbf{R}_s = \sigma_s^2\mathbf{I}_\mathrm{M} \) and \( \mathbf{R}_n = E\{ \mathbf{n}(t) \mathbf{n}(t)^H \} \), and additionally that the transmitted signal and the  distortion are statistically independent, the covariance of the received signal in \eqref{concise receive model} is given by:
\begin{equation}
\begin{split}
\mathbf{R}_y &= E \{ \mathbf{y}(t) \mathbf{y}(t)^H \} \\
&=  \underbrace{\mathbf{a}(\theta) \sigma_s^2 \mathbf{a}(\theta)^H}_{\mathbf{R}_x(\theta)} \odot \underbrace{E \{\mathbf{b}(t) \mathbf{b}(t)^H \}}_{\mathbf{R}_b} + \mathbf{R}_n,
\label{covariance measurement}
\end{split}
\end{equation}
where $\mathbf{R}_x$ is the covariance matrix of the ULA for a single source at angle $\theta$ without rain-induced distortion and $\mathbf{R}_b$ is the distortion covariance matrix.

The elements of the distortion covariance matrix $\mathbf{R}_b$ have the following structure:
\begin{equation}
[\mathbf{R}_b]_{i,j} =
\begin{cases}
E[|b_i(t)|^2]= 2\lambda_{11}, & \text{if } i = j, \\[6pt]
E[b_i(t) b_j^*(t)]=2\alpha_{|i-j|}\lambda_{11}, & \text{if } i \neq j,
\end{cases}
\label{R_phi_structure}
\end{equation}
where $i,j = 1, 2, \dots, M$. The parameter $\alpha_{|i-j|}$ is defined as: $\alpha = \lambda_{13}(d_{|i-j|}) /\ \lambda_{11}$. Consequently, $\mathbf{R}_b$ is a real symmetric Toeplitz matrix.

\section{DoA Estimation Using Covariance Matching}
\label{sec:4}
Generalized least squares (GLS) or covariance matching estimation techniques (COMET) represent a statistical method used to align or match the covariance structures of two datasets or distributions. It aims to adjust the statistical properties of one dataset to match another by aligning their covariance matrices and has been widely applied in array signal processing applications~\cite{kariya2004generalized,OTTERSTEN1998185}.

In this section, we reformulate the DoA estimation problem as a GLS problem. The covariance matrix $\mathbf{R}_y$ can be estimated by taking $T$ measurements and computing the following sample covariance matrix:
\begin{equation}
\mathbf{\hat{R}}_{y} = \frac{1}{T} \sum_{t=1}^{T} \mathbf{y}^{(t)} \, \mathbf{y}^{(t)H}.
\end{equation}

Our goal is to estimate $\theta$ from the covariance matrix estimate $\mathbf{\hat{R}}_y$, considering $\mathbf{R}_b$ as a nuisance parameter. Specifically, we seek to minimize the following cost function:
\begin{equation}
(\hat{\theta},\hat{\mathbf{R}}_b)
= \arg\min_{\theta,\mathbf{R}_b}
\left\lVert
\mathbf{\hat{R}}_{y} \;-\;  \mathbf{R}_x(\theta) \odot \mathbf{R}_b
\right\rVert_{F}^{2}.
\label{cost func}
\end{equation}
In the next section, we will provide methods for solving this optimization problem.

\section{Calibration Algorithm}
\label{sec:5}
For a ULA, the covariance matrix $\mathbf{R}_x$ has a positive semi-definite Hermitian Toeplitz (HT) structure. Since both $\mathbf{R}_x$ and $\mathbf{R}_b$ are HT matrices, their Hadamard product $\mathbf{R}_T=\mathbf{R}_x \odot \mathbf{R}_b$ is also an HT matrix. We adopt an approach that first estimates the combined matrix $\mathbf{R}_T$, and then decouples it by exploiting their structural properties: $\mathbf{R}_x$ consists of unit-modulus complex entries carrying all the phase information, while $\mathbf{R}_b$ is a real-valued Toeplitz matrix. Once $\mathbf{R}_x$ is recovered, subspace-based DoA estimation techniques such as multiple signal classification (MUSIC) can be applied.

The matrix $\mathbf{R}_T$ retains a positive semi-definite HT structure and can thus be fully described by $(2M - 1)$ real-valued parameters. Specifically, it can be decomposed as:
\begin{equation}
\begin{aligned}
\mathbf{R}_T &= \sum_{m=0}^{2M-2} c_m \boldsymbol{\Sigma}_m \\
&= c_0 \mathbf{I}_M + \sum_{m=1}^{M-1} \left( c_m \mathbf{T}_m + c_{m+M-1} \tilde{\mathbf{T}}_m \right),
\label{R_decomposition}
\end{aligned}
\end{equation}

where $\mathbf{T}_m$ is the HT matrix with all zeros except for the entries on the subdiagonals $\pm m$, which contain ones, and similarly, $\tilde{\mathbf{T}}_m$ is the HT matrix with all zeros except for the entries on the subdiagonal $+m$, which contain the imaginary unit $j$, and those on the subdiagonal $-m$, which contain $-j$ \cite{romero2015compression} \cite{yan2024coded}.

Substituting \eqref{R_decomposition} into the cost function \eqref{cost func} and simplifying, we obtain the following new cost function:
\begin{equation}
\hat{\mathbf{c}}
= \underset{\mathbf{c}}{\arg\min} \left\lVert
\mathbf{\hat{R}}_{y} - \sum_{m=0}^{2M-2} c_m \boldsymbol{\Sigma}_m
\right\rVert_{F}^{2}.
\label{cost func3}
\end{equation}
This minimization can be solved using the least squares (LS) method. Constructing the matrix $\mathbf{V}$ by stacking the vectorized basis matrices:
\begin{equation}
\mathbf{V} = \left[\mathrm{vec}(\boldsymbol{\Sigma}_0),\, \mathrm{vec}(\boldsymbol{\Sigma}_1),\, \dots,\, \mathrm{vec}(\boldsymbol{\Sigma}_{2M-2})\right],
\end{equation}
and defining the vector ${\bf c}$ as ${\bf c} = [ c_0, c_1, \dots, c_{2M-2} ]^T$, the problem becomes:
\begin{equation}
\hat{\mathbf{c}} = \underset{\mathbf{c}}{\arg\min} \left\lVert
\mathrm{vec}(\mathbf{\hat{R}}_{y}) - \mathbf{V}\mathbf{c}
\right\rVert_2^2,
\label{alpha_ls}
\end{equation}
with the LS solution given by:
\begin{equation}
\hat{\mathbf{c}} = \left(\mathbf{V}^H \mathbf{V}\right)^{-1} \mathbf{V}^H \, \mathrm{vec}(\mathbf{\hat{R}}_y).
\end{equation}

Once the coefficient vector $\hat{\mathbf{c}}$ is obtained, the estimated matrix $\hat{\mathbf{R}}_T$ can be constructed using the decomposition in~\eqref{R_decomposition}. Next, the decoupling of $\mathbf{R}_x$ and $\mathbf{R}_b$ is performed by separating the phase and magnitude components of $\hat{\mathbf{R}}_T$. The matrix $\mathbf{R}_x$ can be approximately recovered by extracting the phase of each element from the estimated Hadamard product $\hat{\mathbf{R}}_T$, and projecting it onto the unit circle in the complex plane:
\begin{equation}
    \hat{\mathbf{R}}_x = \exp\left(j\, \angle(\hat{\mathbf{R}}_T)\right),
\end{equation}
where $\angle(\cdot)$ denotes the element-wise phase (argument) of a complex matrix.

The matrix $\mathbf{R}_b$ can be estimated by taking the element-wise magnitude of the Hadamard product:
\begin{equation}
    \hat{\mathbf{R}}_b = \left| \hat{\mathbf{R}}_T \right|,
\end{equation}
where $|\cdot|$ denotes the element-wise absolute value operation.

The LS minimization algorithm for the covariance matrix is summarized in Algorithm 1.

\begin{algorithm}[htb]
\caption{Calibration Algorithm}
\begin{algorithmic}[1]
\STATE \textbf{Input} $\mathbf{R}_y$, $\Sigma_0 \dots \Sigma_{2M-1}$.
\STATE $\mathbf{V} = \left[\mathrm{vec}(\boldsymbol{\Sigma}_0),\, \mathrm{vec}(\boldsymbol{\Sigma}_1),\, \dots,\, \mathrm{vec}(\boldsymbol{\Sigma}_{2M-2})\right]$.
\STATE $\hat{\mathbf{c}} = \left(\mathbf{V}^H \mathbf{V}\right)^{-1} \mathbf{V}^H \, \mathrm{vec}(\mathbf{\hat{R}}_y)$.
\STATE $\hat{\mathbf{R}}_T = \sum_{m=0}^{2M-2} \hat{c}_m \boldsymbol{\Sigma}_m$.
\STATE $\hat{\mathbf{R}}_b = \left| \hat{\mathbf{R}}_T \right|$.
\STATE $\hat{\mathbf{R}}_x = \exp\left(j\, \angle(\hat{\mathbf{R}}_T)\right)$.
\label{algorithm1}
\end{algorithmic}
\end{algorithm}

\section{Numerical Results}
\label{sec:6}
In this section, we conduct simulations to evaluate the proposed methods. A ULA with $M = 8$ elements and half-wavelength spacing is assumed, with a single source located at $\theta = 40^\circ$. For the rain-induced distortion model, we assume a rain rate of $50\ \mathrm{mm/hr}$ and a propagation range of $200\ \mathrm{m}$.

\begin{figure}[htbp]
    \centering
    \includegraphics[width=\columnwidth]{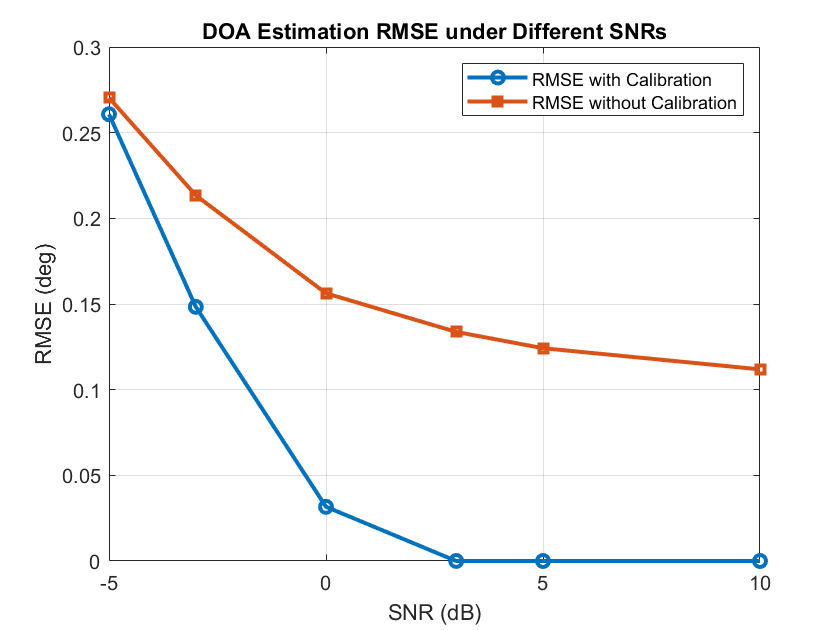}
    \caption{RMSE comparison between calibrated root-MUSIC-based DoA estimation and conventional root-MUSIC without calibration.}
    \label{rmse}
\end{figure}

Figure~\ref{rmse} presents the root mean squared error (RMSE) performance of the DoA estimation methods. It compares the proposed calibration-enhanced root-MUSIC approach with the conventional root-MUSIC method without calibration. The results demonstrate that the proposed calibration significantly improves angle estimation accuracy.

\begin{figure}[htbp]
    \centering
    \includegraphics[width=\columnwidth]{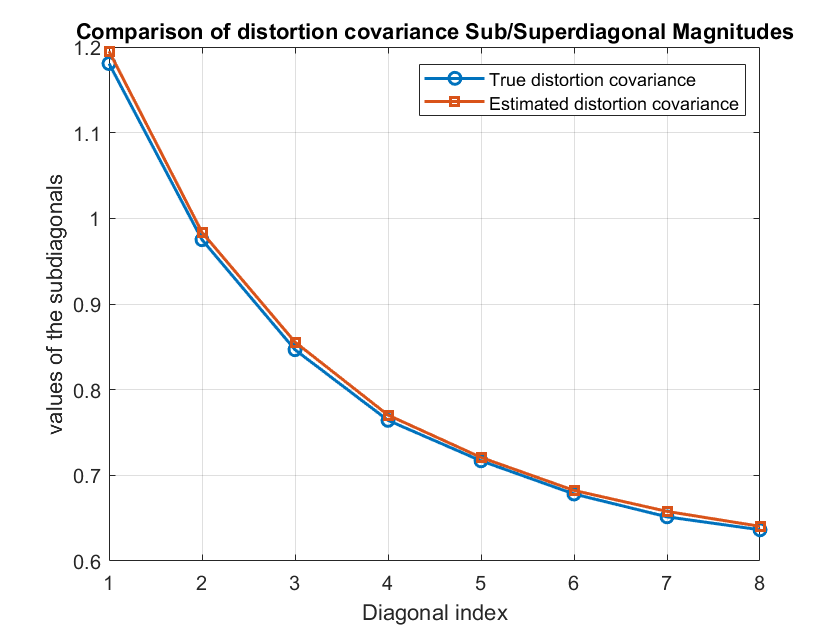}
    \caption{Subdiagonal-wise comparison between the estimated and true distortion covariance matrices (results obtained at $\mathrm{SNR}=20\,\mathrm{dB}$).}
    \label{distortioncov}
\end{figure}

Figure~\ref{distortioncov} presents a comparison of the subdiagonal values in the Toeplitz structure between the estimated distortion covariance matrix and the true covariance matrix under $\mathrm{SNR}=20\,\mathrm{dB}$. The strong agreement across subdiagonals indicates that the proposed estimation method effectively captures the underlying distortion statistics.

\begin{figure}[htbp]
    \centering
    \begin{subfigure}{\columnwidth}
        \centering
        \includegraphics[width=\linewidth]{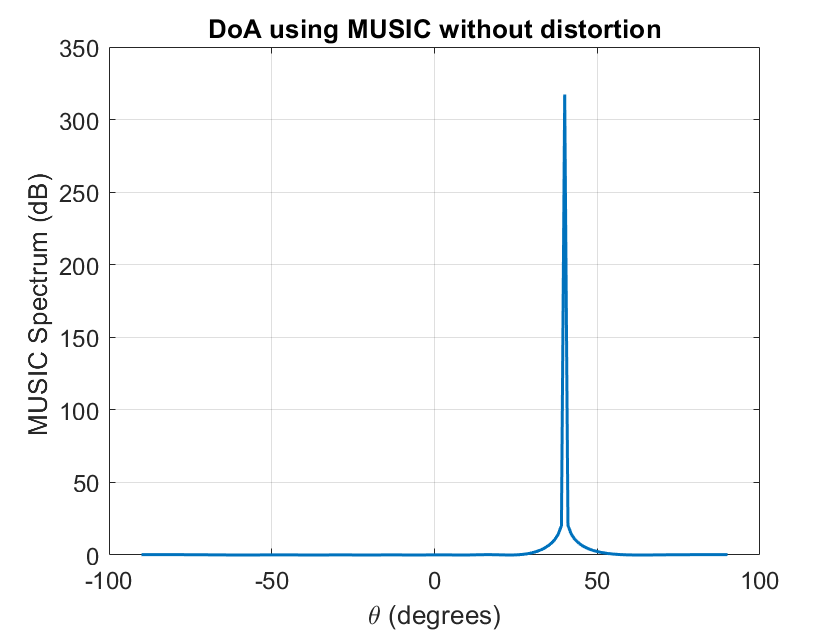}
        \caption{No rain}
        \label{fig:MUSIC_norain}
    \end{subfigure}
    \hfill
    \begin{subfigure}{\columnwidth}
        \centering
        \includegraphics[width=\linewidth]{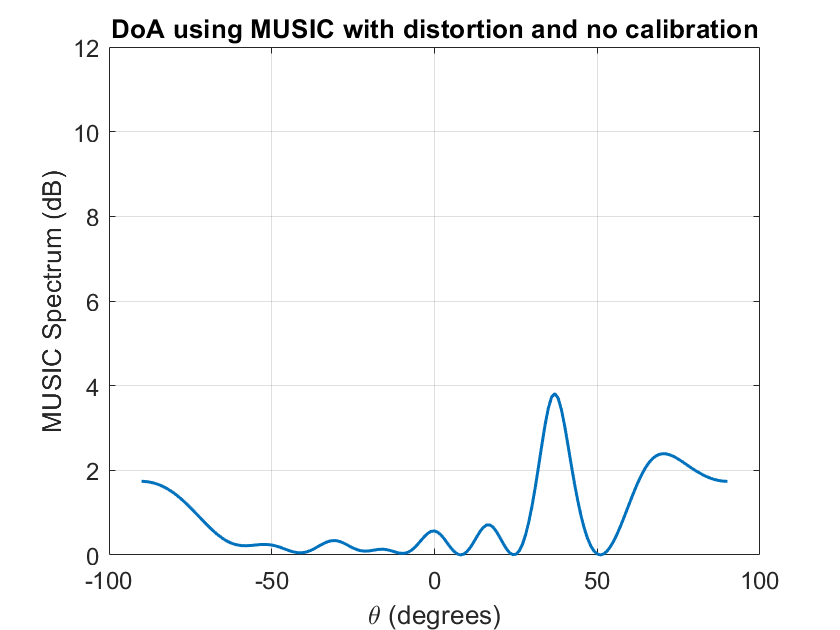}
        \caption{Rain, no calibration}
        \label{fig:MUSIC_rain}
    \end{subfigure}
    \hfill
    \begin{subfigure}{\columnwidth}
        \centering
        \includegraphics[width=\linewidth]{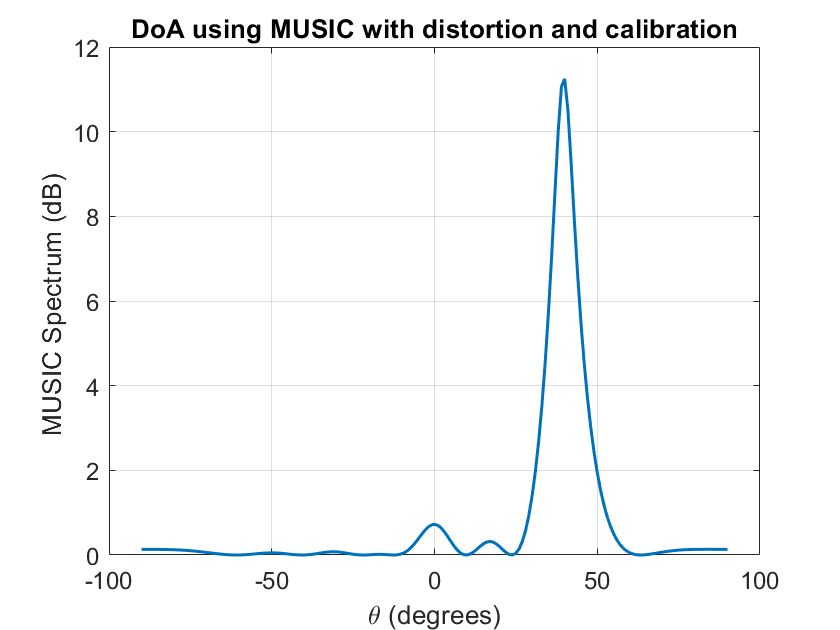}
        \caption{Rain, with calibration}
        \label{fig:MUSIC_calibration}
    \end{subfigure}
    \caption{MUSIC spectra in dB scale under three conditions: (a) no rain distortion, (b) rain distortion without calibration, and (c) rain distortion with calibration.}
    \label{fig:MUSIC_all}
\end{figure}

Figure~\ref{fig:MUSIC_all} shows the MUSIC spectra under three different scenarios: (a) without rain distortion, (b) with rain distortion but without calibration, and (c) with rain distortion and calibration. It is evident that the proposed calibration method effectively mitigates the impact of rain-induced phase errors, resulting in a sharper spectral peak at the true angle and enhancing the resolution of DoA estimation.

\section{Conclusions}
\label{sec:7}
We have proposed a ULA model incorporating rain-induced phase aberrations and have applied a GLS approach to enhance DoA estimation accuracy. Simulation results demonstrate effective mitigation of rain-induced distortions and improved angle estimation performance. The proposed method can be directly applied to other types of phase noise characterized by symmetric pdfs.

While this is an initial study, future research directions include extending the approach to multiple sources, 3D angle estimation, and refining the model to account for unknown propagation gains, among other factors.

\bibliographystyle{IEEEtran}
\bibliography{ref}

\end{document}